\documentclass[prd,amsmath,floats,amssymb, floatfix,superscriptaddress,nofootinbib,onecolumn]{revtex4} 

\usepackage[plainpages=false, colorlinks=true, anchorcolor=blue, linkcolor=blue, citecolor=blue, bookmarks=false]{hyperref}
\usepackage[T1]{fontenc}
\usepackage{graphicx}
\usepackage{dcolumn}
\usepackage{bm}
\usepackage{longtable}
\usepackage{adjustbox}
\newcommand{\rthis}[1]{\textcolor{black}{#1}}
\usepackage{float}
\usepackage[caption=false]{subfig}
\usepackage[paperwidth=210mm,paperheight=297mm,centering,hmargin=2cm,vmargin=2.5cm]{geometry}

\begin{document}
\include{notations}
\preprint{APS/123-QED}

\title{ A pilot search  for MeV gamma-ray emission from five galaxy clusters using  archival COMPTEL data}

\author{Siddhant Manna}
 \altaffiliation{Email:ph22resch11006@iith.ac.in}
\author{Shantanu Desai}
 \altaffiliation{Email:shntn05@gmail.com}
\affiliation{
 Department of Physics, IIT Hyderabad Kandi, Telangana 502284,  India}





\begin{abstract}
We search for MeV gamma-ray emission between 0.75-30 MeV from five galaxy clusters, viz. Coma, VIRGO, SPT-CL J2012-5649, Bullet, and El Gordo, using archival data from the COMPTEL telescope. For this purpose we use three search templates: point source, radial disk and radial Gaussian. We do not detect any signals from Coma,  SPT-CL J2012-5649, Bullet and El Gordo clusters with the 95\% c.l. photon energy flux limit  $\sim 10^{-10} ~\rm{erg /cm^2/s} $. 
For VIRGO, we detect a non-zero signal between 0.75 to 1.50 MeV  having  marginal significance of  about $2.5\sigma$,   with the observed energy flux equal to  $ \sim 10^{-9}~\rm{ergs/cm^2/s}$. However, we do not confirm  the previously reported evidence in literature for a gamma-ray line  from Coma and VIRGO clusters between 5-7 MeV.

\end{abstract}


\maketitle
\section{\label{sec:level1}Introduction\protect}
Galaxy clusters  are   the largest gravitationally bound and virialized structures in the Universe and act as  wonderful laboratories for Cosmology, galaxy evolution~\cite{Kravtsov2012, Allen2011, Vikhlininrev} and fundamental Physics~\cite{Bohringer16,Desai18, Boraalpha, BoraDesaiCDDR, Mendonca,BoraDM}. They  have been observed over a broad range of the electromagnetic spectrum   from  radio waves~\cite{Feretti} to hard X-rays  with energies up to 30 keV~\cite{Wik14,Rojas}. In the last two decades a large number of new galaxy clusters have been discovered through dedicated surveys at optical, X-ray, and millimeter wavelengths. At higher energies, the detection of gamma-rays from galaxy clusters (at GeV energies) is still a hotly debated issue, with only the Coma cluster showing  evidence for gamma-ray emission from the intra-cluster medium~\cite{Reisscoma,Xi18,Baghmanyan22}.  Gamma-rays could be produced in galaxy clusters due to hadronic interactions or inverse Compton scattering  processes in the intra-cluster medium~\cite{Christoph}. Another possible source of gamma-rays  in clusters could be from  dark matter annihilations if the dark matter is a WIMP~\cite{Ullio06}. All possible mechanisms of gamma-ray production in clusters have been recently reviewed in ~\cite{Manna23}.

Most recently, we carried out a systematic search for gamma-ray emission between 1-300 GeV  using 15 years of Fermi-LAT data.~\cite{Manna23}, from a sample of 300 galaxy clusters selected from the South Pole Telescope 2500 square degree cluster survey~\cite{Bleem15,Bocquet19}, whose redshifts were estimated from optical/NIR follow-ups~\cite{Desai12,Song}. The clusters in the SPT-SZ survey have been detected using the Sunyaev-Zeldovich effect~\cite{SZ} and hence represent a mass limited sample. From this catalog, we were able to detect  gamma-ray emission from one galaxy  cluster with significance $ > 5\sigma$, viz. SPT-CLJ 2012-5649 with six other clusters having significance between $3-5\sigma$~\cite{Manna23}.  However  we could not discern if this emission is from the intra-cluster medium or because of radio galaxies in the  cluster.

However, there is a paucity of studies on clusters in the soft gamma-ray energy band  between  hard X-rays and gamma-rays at MeV energies . One reason is that there has only been one space-based telescope  that has imaged the sky in this energy range  from 0.75-30 MeV, which is the Imaging Compton Telescope (COMPTEL). COMPTEL was one of the three telescopes  onboard the Compton Gamma Ray Observatory (CGRO), which operated from 1991-2000~\cite{Comptel93}. No other space-based gamma-ray telescope launched within the last two decades has covered the above energy range.
The Fermi-LAT and AGILE telescopes  are sensitive at higher energies ($\geq$ 30 MeV), whereas  INTEGRAL is sensitive up to 10 MeV. Therefore, there has only been one paper in literature which has searched for gamma-ray emission from galaxy clusters at energies between between 0.75-30 MeV  using COMPTEL~\cite{Iyudin} (I04, hereafter). This work  looked for MeV emission from galaxy clusters near the North Galactic pole.
I04 found evidence for a gamma-ray line between 5.6-7.6 MeV  with statistical significance of around   $5\sigma$  from Coma cluster as well as from two elliptical galaxies M87  and M49 in the   VIRGO cluster, which they argued  is due to the  de-excitation gamma-ray line emission from $^{16}\rm{O}$, which could be produced from spallation interactions of cosmic rays with ions in the intra-cluster medium.
The observed  $^{16}\rm{O}$ $\gamma$-ray line flux  was equal to $1.1 \times 10^{-5}$ photons/$cm^{2}$/sec and $1.1 \times 10^{-5}$ photons/$cm^{2}$/sec, from VIRGO and Coma, respectively. \rthis{We also note that M87 is a FR-I radio galaxy with a  misaligned jet and has  been  detected at very high energies ($E> 100$ GeV) by Fermi-LAT, H.E.S.S., MAGIC, and VERITAS~\cite{Molero}. }

Although the  COMPTEL data is still archived
at NASA's HEASARC, the legacy software used to process the COMPTEL data was decommissioned shortly after the end of CGRO mission and was not easily accessible to the wider scientific community. Therefore, there were only a handful of results from the COMPTEL mission in the last twenty odd years~(\cite{Strong} and references therein). Recently, a new plugin as well as custom Python scripts  have been added to the widely used GammaLib and ctools libraries~\cite{gammalib} to analyze the  COMPTEL data,  facilitating a seamless scientific analysis~\cite{Knod}. Given that the energy range probed by COMPTEL is an uncharted territory with no approved experiment in the near future probing this energy range, it behooves us to  analyze  this legacy data to look for gamma-ray emissions from  galaxy clusters. 

Given some of the intriguing hints for GeV gamma-ray emission from galaxy clusters~\cite{Manna23}, it would be interesting to look for detections at MeV energies from the same clusters. Therefore, as a pilot search, we analyze the archival COMPTEL data to look for detections at MeV energies . The targets we consider for our analysis include VIRGO, Coma, SPT-CLJ 2012-5649, Bullet (SPT-CL J0658-5556), and El Gordo (SPT-CL J0102-4915) clusters. As mentioned above, I04 had detected MeV emission from both   Coma and the VIRGO clusters. Most recently, we detected $5\sigma$ emission from SPT-CL J2012-5649 (also known as Abell 3667) between 1-10 GeV using the Fermi-LAT telescope. Therefore, it would be interesting to check if the signal seen in SPT-CL J2012-5649 at GeV energies persists at lower energies. Both the Bullet and  El Gordo clusters are massive merging clusters,  which have proved to be a very good laboratories  for testing  $\Lambda$CDM and modified gravity theories~\cite{Clowe06,Kroupa}.  Therefore they would also be very good targets for gamma-ray searches.

This manuscript is structured as follows. In Sect.~\ref{sec:level2}, we describe the COMPTEL dataset used for analysis.   Our results can be found in Sect.~\ref{sec:results}. Our conclusions can be found in Sect.~\ref{sec:conclusions}. 



\section{Data Analysis}
\label{sec:level2}
The COMPTEL telescope, onboard the Compton Gamma Ray Observatory (CGRO) launched in 1991, was a pioneering instrument that revolutionized our understanding of the high-energy universe. It employed a unique imaging technique called Compton scattering, where gamma rays interact with electrons within the detector, changing their direction. By analyzing these scattered photons, it could reconstruct the location and energy of the original gamma-ray source. The telescope field of view was about one steradian. It was sensitive to gamma-rays between 0.75- 30 MeV, where  the energy and angular resolution ranged between 5-8\% and $(1.7-4.4)^{\circ}$, respectively depending on the photon energy. During its 9.7 years of operation, COMPTEL did about 340 distinct pointings (with a duration of two weeks), where each pointing had a field of view radius of around $30^{\circ}$~\cite{Strong}. More details on the observing specifications, detector calibration,  and performance of COMPTEL can be found in ~\cite{Comptel93}. A large number of astrophysical sources have been detected by COMPTEL such as pulsars, AGNs, X-ray binaries, gamma-ray bursts, solar flares, $^{26}$Al, supernova remnants, extragalactic diffuse gamma-ray background, etc~\cite{Comptel93,Comptel00}.

We comprehensively analysed the COMPTEL data to investigate the potential presence of MeV gamma-ray emission from five  galaxy clusters: the Coma Cluster, SPT-CL J2012-5649 Cluster, the Bullet Cluster,the El Gordo Cluster and the Virgo Cluster. We carried out  a systematic search within a $10{^\circ}$ radius surrounding each target cluster utilizing the revamped COMPTEL analysis framework implemented in the ctools software package~\cite{Knod}. We employed the \texttt{comobsselect} tool to extract the pertinent viewing periods for the clusters under investigation, applying precise filtering based on their coordinates and a specified radius. The selection process effectively removes unwanted background noise. Using the \texttt{comobsbin} tool, we implemented energy binning to optimize  the signal-to-noise ratio. We used 16 logarithmically-spaced energy bins spanning the 0.75-30 MeV range. With this spacing, we could detect the  gamma-ray lines which have been reported in I04.
For SPT-CL J2012-5649, El Gordo, VIRGO and Bullet Cluster, we combined the data from  two or more distinct viewing periods using the \texttt{comobsadd} tool with 80 bins each in the $\chi$ and $\psi$ directions, where $\chi$ and $\psi$ are the Compton scattering directions. Unlike the other clusters, for Coma cluster, we got data from only one viewing period in our specified radius ($r=10{^\circ}$). The effective exposure for all our clusters can be found in  Table~\ref{tab:ts_values}.

Our search analysis was done using three different templates, viz. point source, radial disk and radial Gaussian templates for our analysis. We used the same radial model definitions used in Fermi-LAT analyses~\cite{Wood2017}~\footnote{available at  \url{https://fermi.gsfc.nasa.gov/ssc/data/analysis/scitools/xml_model_defs.html}}.
For the spectral analysis, we utilized the \texttt{comobsmodel} tool to fit source emission models to the COMPTEL data. 
We then employed the \texttt{comlixfit} tool to perform maximum likelihood fitting of the COMPTEL data utilizing the iterative \texttt{SRCLIX} algorithm. Furthermore, the instrumental background Model was modelled using the \texttt{BGDLIXE} background computation model.  By accounting for the instrumental background components, \texttt{BGDLIXE} enables a more accurate and reliable characterization of the true MeV gamma-ray signal originating from the target galaxy clusters. We leveraged the \texttt{comlixmap} tool to construct Test Statistics (TS) maps (cf. Eq.~\ref{eq:TS}) by applying the \texttt{SRCLIX} on every test source position. The test statistic provides a quantitative indication of the evidence against a null hypothesis. Between a model with both signal and background components ($M_s+M_b$) and a model with only the background ($M_b$), the test statistic (TS) can be expressed as~\cite{Mattox1996}:
\begin{equation}
TS = 2 \ln L(M_s + M_b) - 2 \ln L(M_b),
\label{eq:TS}
\end{equation}
where $\ln L(M_s + M_b)$  denotes  the likelihood when both the source model ($M_s$) and the background model ($M_b$) are jointly fitted to the data. It reflects how well the combined model aligns with the observed data, while $\ln L(M_b)$ corresponds to the log-likelihood, when only the background model is considered. Crucially, under the assumption that the background model adequately captures the data, the TS follows a $\chi^{2}$ distribution with $n$ degrees of freedom, where $n$ denotes the number of free parameters within the source model component.  The same statistics is also used in IceCube and Fermi-LAT analysis~\cite{Pasumarti,Manna23}. 

For each cluster we also constructed the Spectral Energy Distribution (SED) plots using the \texttt{csspec} tool and obtained upper limits (in case of null results) with the \texttt{ctulimit} tool. These SED plots, constructed within the energy range of 0.75-30 MeV using 16 logarithmic energy bins and the \texttt{BINS} spectrum generation method, visually represent the upper limits on the flux density of potential gamma-ray sources within the clusters. The upper limits for the differential photon flux were determined using a reference energy of 30 MeV, chosen to maximize the sensitivity to potential spectral variations within the MeV range. We used \texttt{BINS} method to run the \texttt{csspec} script, the spectral model is replaced by a binned function that fits all data.
We calculate integral photon and energy flux upper limits over the full energy range of 0.75 MeV to 30 MeV, encompassing the analyzed spectrum to provide  constraints for clusters with null detections. The energy flux signifies the cumulative  photon energy deposited within a specified energy range.

\section{Results}
\label{sec:results}
\begin{figure}[h]
  \centering
  \includegraphics[width=0.8\textwidth]{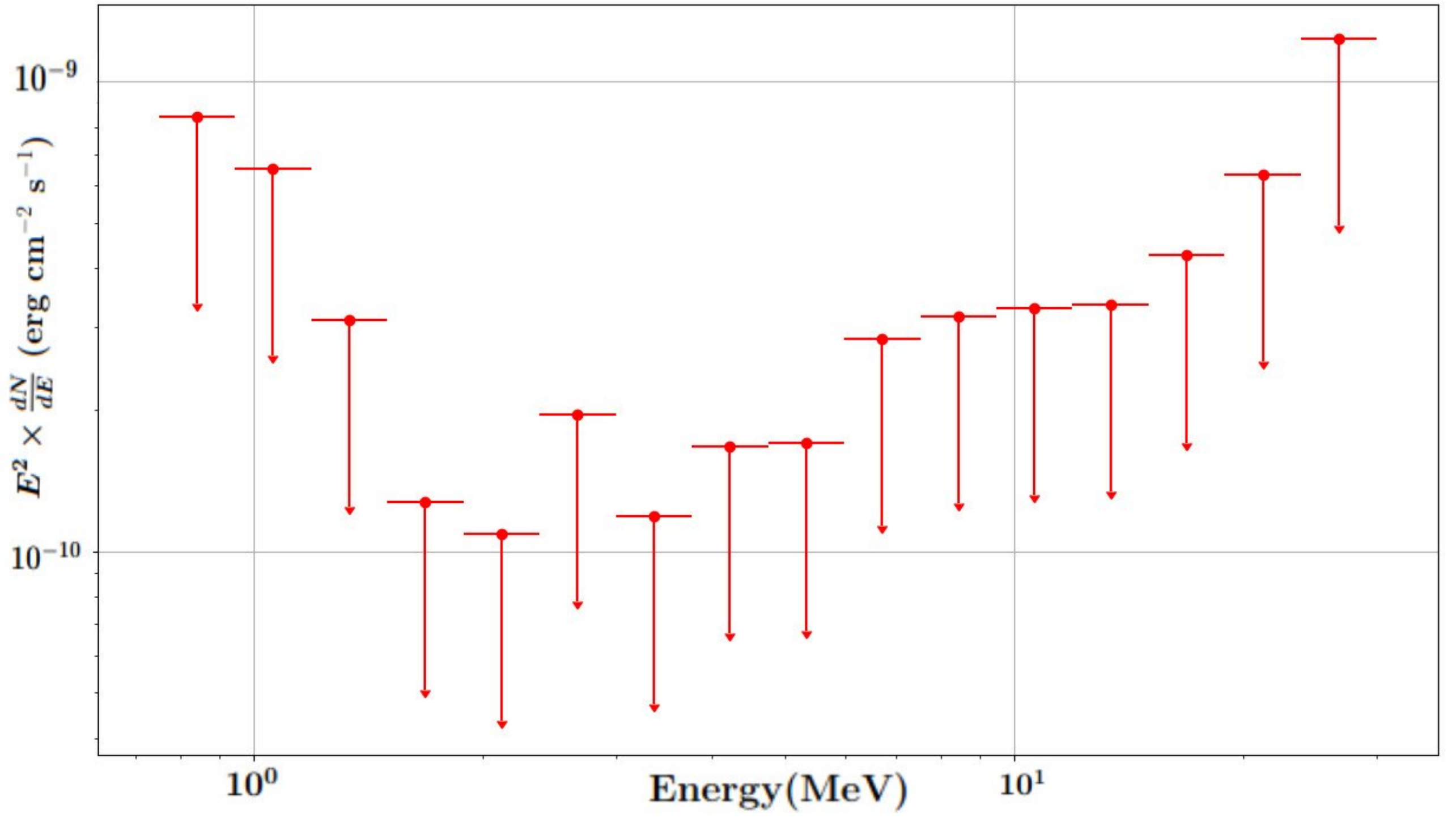}
  \caption{SED plot for Coma Cluster using a point source template
}
  \label{fig:Figure1}
\end{figure}

\begin{figure}[h]
  \centering
  \includegraphics[width=0.8\textwidth]{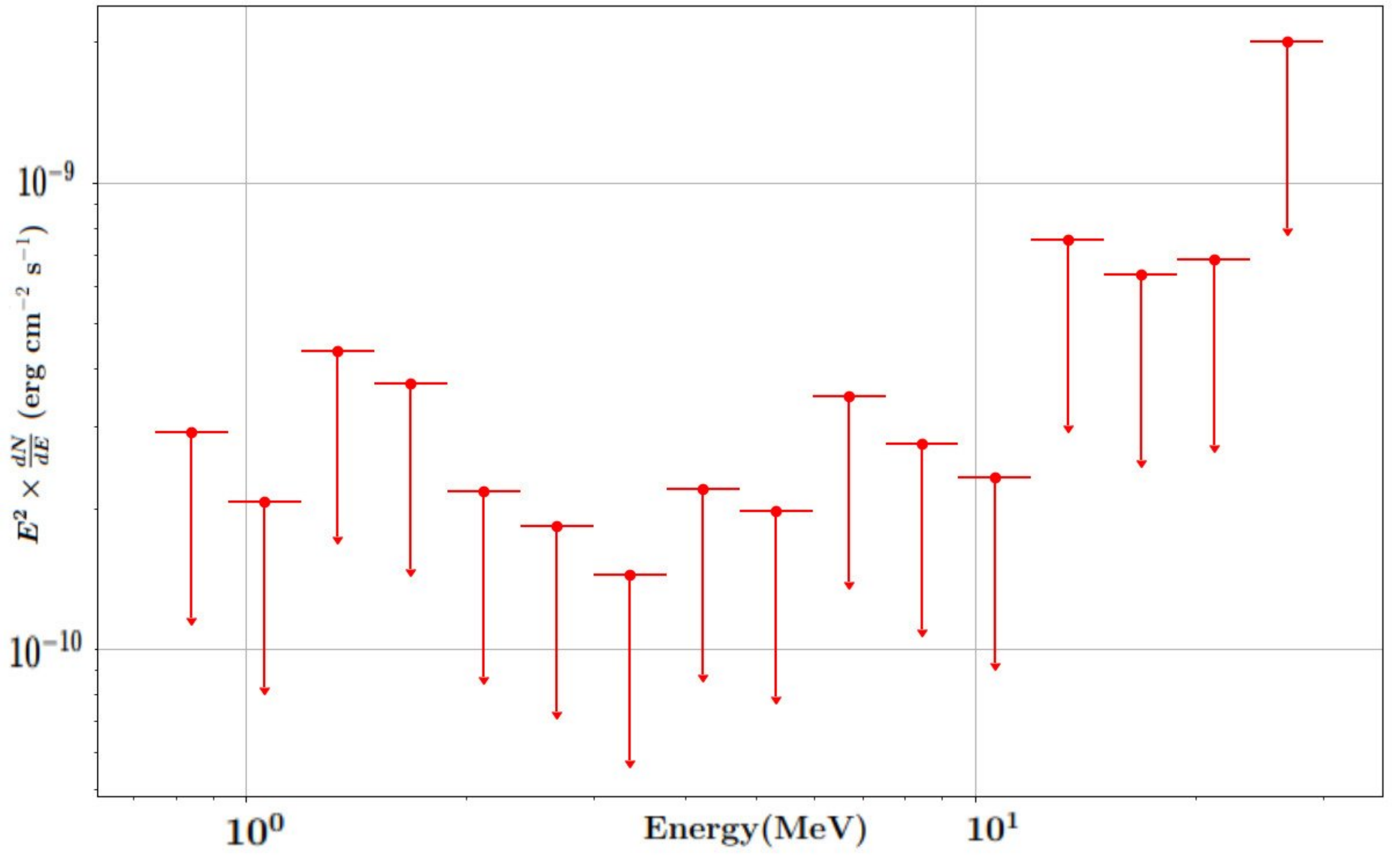}
  \caption{SED plot for  SPT-CL J2012-5649 Cluster using a point source template.}
  \label{fig:Figure2}
\end{figure}

\begin{figure}[h]
  \centering
  \includegraphics[width=0.8\textwidth]{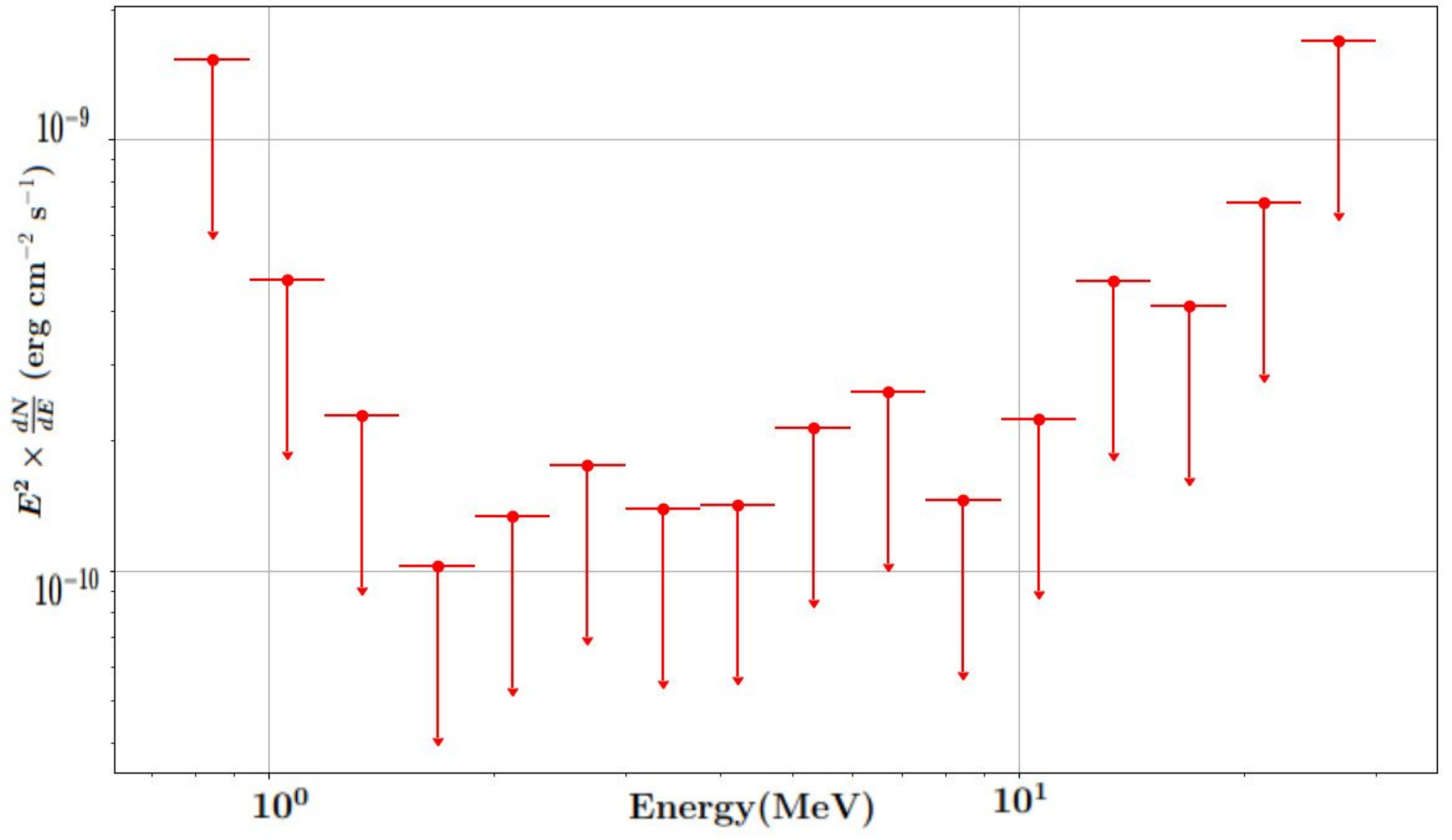}
  \caption{SED plot for the Bullet Cluster using a point source template.
}
  \label{fig:Figure3}
\end{figure}

\begin{figure}[h]
  \centering
  \includegraphics[width=0.8\textwidth]{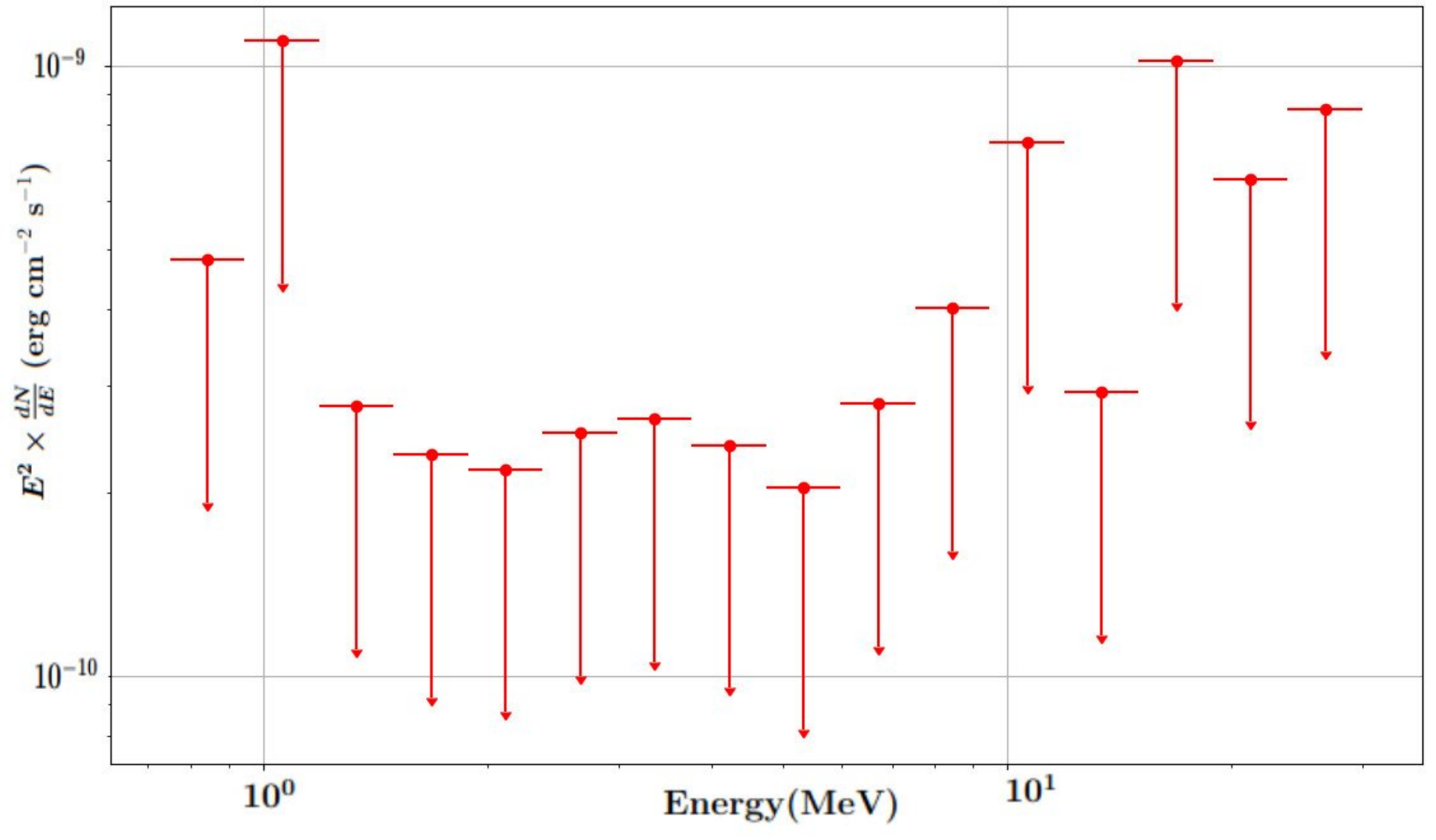}
  \caption{SED plot for El Gordo  using a point source template
}
  \label{fig:Figure4}
\end{figure}

We now report the results from our searches from the five clusters  between the energy range of 0.75-30 MeV, spanning the full COMPTEL observing period from 1991 to 2000.
Table \ref{tab:ts_values} summarizes  the maximum TS  value for each cluster for all the three templates  along with the effective exposure. For Coma,  SPT-CL J2012-5649, Bullet, and El Gordo, the TS values are less than 3.0. For VIRGO we see TS values greater than 3.0 for all the three input templates, with the largest value  of 5.6 obtained for the radial disk template. Consequently, the corresponding SED plots only show upper limits for all the clusters  except VIRGO. For illustration, we show the  corresponding SED plots for a point source template  for Coma, SPT-CL J2012-5649, Bullet, and El Gordo clusters in 
Figures \ref{fig:Figure1}, \ref{fig:Figure2}, \ref{fig:Figure3}, \ref{fig:Figure4}, respectively. The upper limits for the other templates are of similar values.
For the VIRGO cluster, we show the SED plots for all the three templates. We obtain a non-zero flux in the first three energy bins between $0.75$ to $1.50$ MeV, having  marginal significance between $(2.5-2.8)\sigma$ and null detections at  higher energies. \rthis{For the VIRGO cluster, we first used the fixed position of RA=$187.71^{\circ}$ and Dec=$12.39^{\circ}$. We also redid the fit by keeping the position a free parameter and we get a best position consistent with our initial position and with the same SED as before. For the Coma as well as VIRGO cluster we also redid our analyses with different binnings including the one used in I04, but it does not change our results. }
Therefore, we cannot reproduce the results for line emission from the Coma cluster reported in I04. Although we detect a signal from the VIRGO cluster, it is seen at a lower energy than what was found in I04.  

We summarize our results for all the clusters in Table \ref{tab:flux_limits}. We report the 95\% confidence level upper limits for the differential photon flux (at 30 MeV), the integral photon flux, and photon  energy flux. For Coma, SPT-CL J2012-5649, Bullet, and El Gordo, we report the differential flux upper limit at 30 MeV, along with integral flux and energy flux upper limits. 
For Coma cluster, the integral photon flux limit  we obtain of about $(1.4-1.5) \times 10^{-5}~\rm{cm^{-2} s^{-1}}$ is comparable to the reported flux in I04.
All the remaining clusters with null results also show similar values for the differential and integral photon flux as well as the energy flux.
For the VIRGO cluster, we report the aforementioned measured quantities along with $1\sigma$ error bars, which can be found in the last three rows of Table~\ref{tab:flux_limits}. For the point source template, the total energy flux equals $(0.57 \pm 0.25) \times 10^{-9}$ erg/cm$^2$/s. When we use the  radial disk and radial Gaussian templates, we get total photon energy fluxes of $(1.25 \pm 0.48) \times 10^{-9}$ erg/cm$^2$/s and $(0.70 \pm 0.28) \times 10^{-9}$ erg/cm$^2$/s respectively. 
Once again, the photon flux is comparable to the $^{16} O$ line flux reported in I04, albeit measured  at lower energies. \rthis{We note that the AGN in M87 could be the cause of the observed MeV emission,s ince M87 has been detected at very high energies.}

\begin{table}[htbp]
\caption{ The maximum TS values for the five  galaxy clusters in the energy range between 0.75-30 MeV,  along with effective exposure for all the three search templates employed.}
\label{tab:ts_values}
\centering
\begin{tabular}{|l|c|c|c|}
\hline
\textbf{Cluster} & \textbf{Template} & \textbf{TS Values}  & \textbf{Exposure} \newline \textbf{(cm$^2$/s)}\\
\hline
             & Point & 0.02 & - \\
Coma & Radial Disk & 0.06 & $3.78\times 10^{9}$ \\
             & Radial Gaussian & 0.01 &  \\
                  & Point & 1.44 & - \\
SPT-CL J2012-5649 & Radial Disk & 2.74 & $0.84\times 10^{9}$ \\
                  & Radial Gaussian & 1.9 &  \\
               & Point & 0 & - \\
Bullet  & Radial Disk & 0 & $3.94\times 10^{9}$ \\
               & Radial Gaussian & 0 &   \\
                 & Point & 1.12 & - \\
El Gordo & Radial Disk & 2.40 & $1.74\times 10^{9}$ \\
                 & Radial Gaussian & 1.00 &  \\
                 & Point  & 3.0  & - \\
VIRGO   & Radial Disk  & 5.6 & $1.12\times 10^{9}$ \\
                 & Radial Gaussian &  3.6 &  \\
\hline
\end{tabular}
\end{table}

\begin{table*}[htbp]
\caption{Results for  MeV gamma-ray emission from five galaxy clusters searched using  COMPTEL data. For the first four clusters, we report  $95\%$ c.l. upper limits for
differential flux (at 30 MeV), integral Flux and energy Flux 
and are shown with $<$. For the VIRGO cluster, we report the observed values for the same along with $1\sigma$ error bars.}
\label{tab:flux_limits}
\centering
\begin{tabular}{|p{3cm}|p{3cm}|p{3cm}|p{3cm}|p{3cm}|l|c|c|c|c|c|}
\hline
\textbf{Cluster} & \textbf{Template}  & \textbf{Differential Flux at 30 MeV} \newline \textbf{(ph/cm$^2$/s/MeV)} & \textbf{Integral Flux} \newline \textbf{(ph/cm$^2$/s)} & \textbf{Energy Flux} \newline \textbf{(erg/cm$^2$/s)} \\
\hline
    & Point & $<1.18\times 10^{-8}$ & $<1.39\times 10^{-5}$ & $<6.30\times 10^{-11}$ \\
Coma  & Radial Disk  & $<1.31\times 10^{-8}$ & $<1.54\times 10^{-5}$ & $<6.98\times 10^{-11}$ \\
  & Radial Gaussian & $<1.21\times 10^{-8}$  & $<1.42\times 10^{-5}$ & $<6.44\times 10^{-11}$ \\
 & Point & $<1.35\times 10^{-8}$ & $<1.59\times 10^{-5}$ & $<7.21\times 10^{-11}$ \\
SPT-CL J2012-5649 & Radial Disk & $<1.70\times 10^{-8}$ & $<1.98\times 10^{-5}$ & $<9.03\times 10^{-11}$ \\
 & Radial Gaussian & $<1.39\times 10^{-8}$ & $<1.63\times 10^{-5}$ & $<7.40\times 10^{-11}$ \\
 & Point & $<1.13\times 10^{-8}$ & $<1.33\times 10^{-5}$ & $<6.03\times 10^{-11}$ \\
Bullet  & Radial Disk & $<1.32\times 10^{-8}$ & $<1.55\times 10^{-5}$ & $<7.04\times 10^{-11}$ \\
& Radial Gaussian  & $<1.15\times 10^{-8}$ & $<1.35\times 10^{-5}$ & $<6.13\times 10^{-11}$ \\
 & Point  & $<1.84\times 10^{-8}$  & $<2.15\times 10^{-5}$ & $<9.80\times 10^{-11}$ \\
El Gordo & Radial Disk  & $<1.97\times 10^{-8}$  & $<2.33\times 10^{-5}$ & $<9.94\times 10^{-11}$ \\
& Radial Gaussian  & $<1.80\times 10^{-8}$  & $<2.11\times 10^{-5}$ & $<9.58\times 10^{-11}$ \\
 & Point  & $(0.35 \pm 0.21) \times 10^{-3}$  & $(0.22 \pm 0.09) \times 10^{-5}$ & $(0.57 \pm 0.25) \times 10^{-9}$ \\
VIRGO  & Radial Disk & $(0.75 \pm 0.40) \times 10^{-3}$ & $(0.48 \pm 0.17) \times 10^{-5}$ & $(1.25 \pm 0.48) \times 10^{-9}$  \\
& Radial Gaussian  & $(0.42 \pm 0.24)  \times 10^{-3}$ & $(0.27 \pm 0.09) \times 10^{-5}$ & $(0.70 \pm 0.28) \times 10^{-9}$ \\
\hline
\end{tabular}
\end{table*}

\begin{figure}[h]
  \centering
  \includegraphics[width=0.8\textwidth]{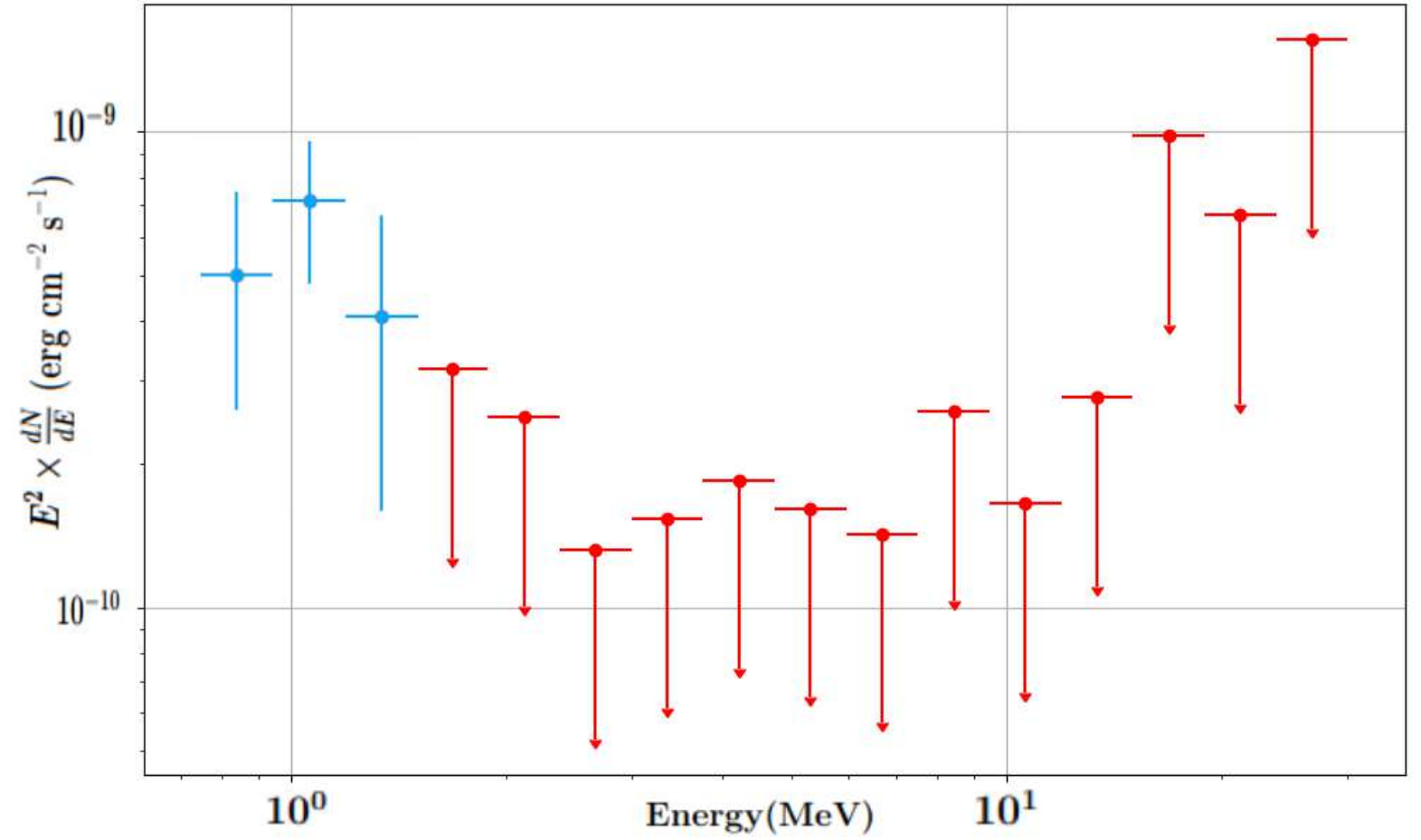}
  \caption{SED plot for the  VIRGO Cluster using point source template. The first three data points in blue show the measured non-zero flux, whereas those in  red  color at higher energies represent  upper limits.The data points shown are the center of energy bins.}
  \label{fig:Figure5}
\end{figure}
\begin{figure}[h]
  \centering
  \includegraphics[width=0.8\textwidth]{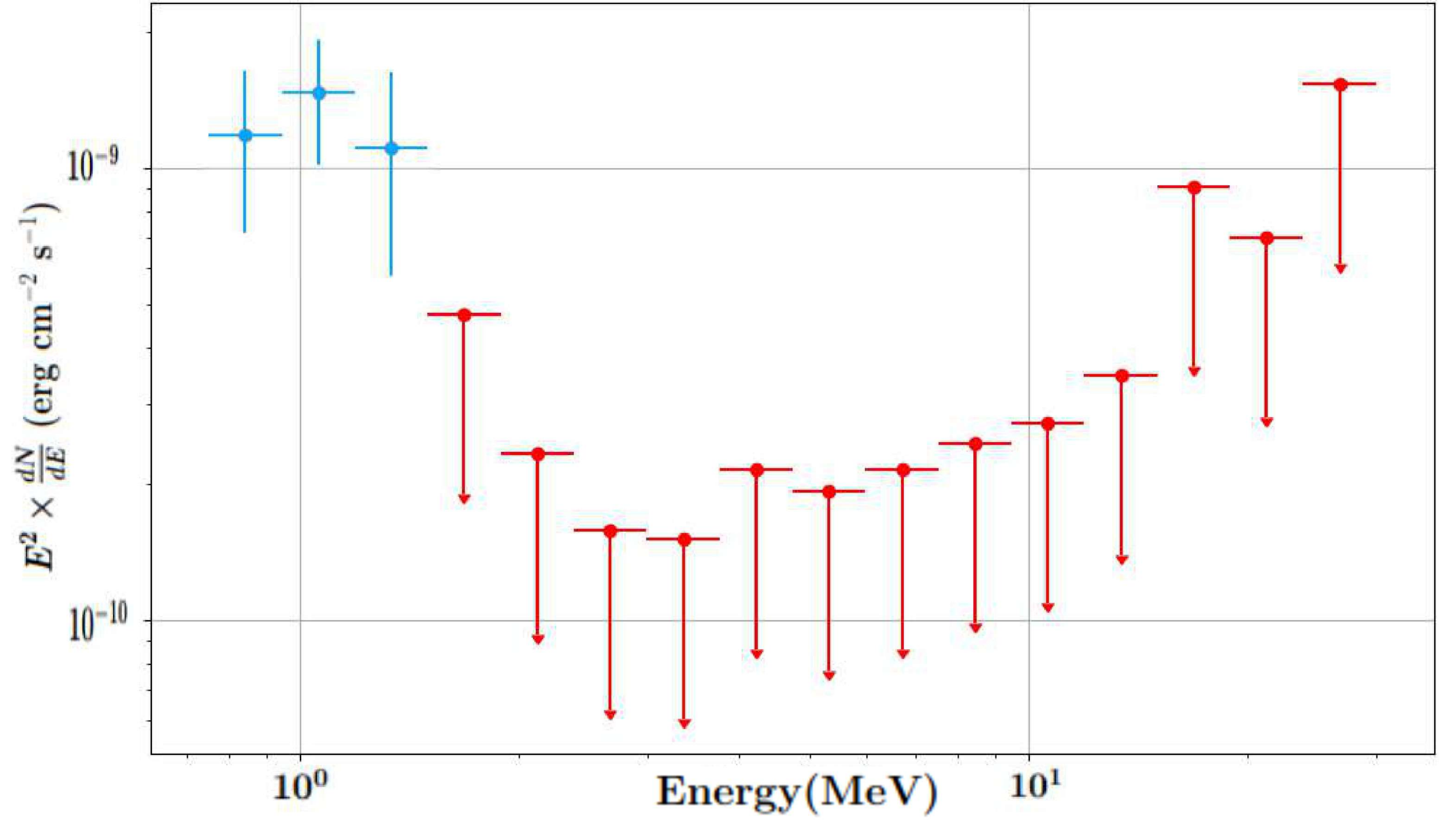}
  \caption{Same as \ref{fig:Figure5} but the SED plot is obtained  using the radial disk template.}
  \label{fig:Figure6}
\end{figure}
\begin{figure}[h]
  \centering
  \includegraphics[width=0.8\textwidth]{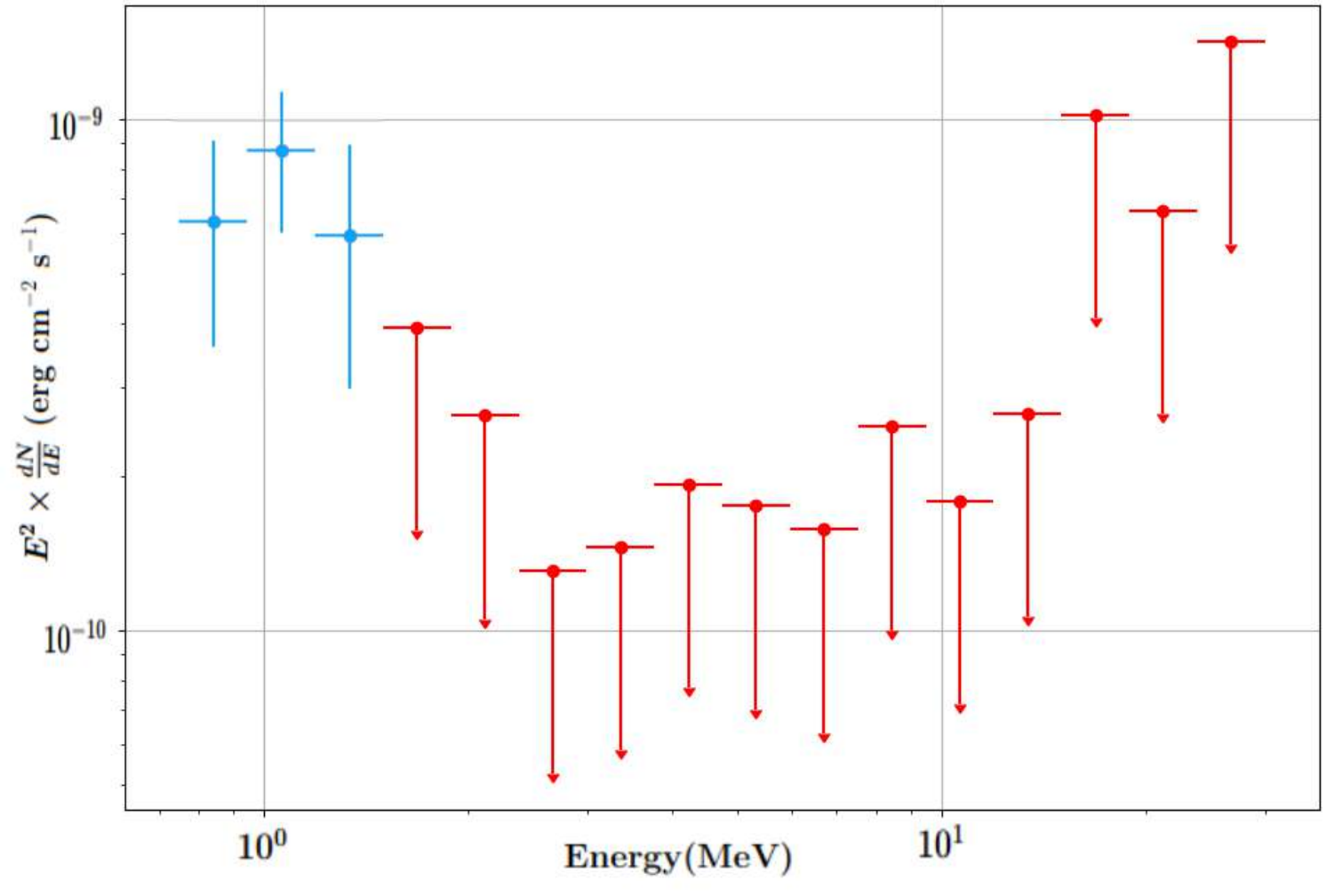}
  \caption{Same as \ref{fig:Figure5} but the SED plot is  obtained using the radial Gaussian template.}
  \label{fig:Figure7}
\end{figure}

\section{\label{sec:conclusions}Conclusions\protect}

There has only been one previous search (reported in I04) for gamma-ray emission from galaxy clusters in the energy range  between 0.75-30 MeV using the COMPTEL telescope. I04 had found a gamma-ray line between 5-7 MeV from two galaxy clusters namely, VIRGO and Coma. Although, there have been a plethora of sources detected  at GeV  energies from the Fermi-LAT satellite, no other systematic search for emission from galaxy clusters has been carried in the MeV energy  region. The main reason is that there is no other telescope apart from COMPTEL, which has mapped the universe in MeV gamma-rays. Even the COMPTEL telescope was decommissioned in 2000 and the software used to analyze the data was  available only to  the core institutes involved in the COMPTEL collaboration.

In the last two years, the software used to analyze the archival COMPTEL data has been revamped and made publicly available, facilitating a seamless analysis of this data~\cite{Knod}. We leverage on this to carry out a pilot search for MeV gamma-ray  emission using COMPTEL from five galaxy clusters, namely Coma, VIRGO, Bullet, El Gordo and SPT-CL J2012-5649. We used three search templates for the same: point source, radial disk and radial Gaussian. Our results are summarized in Table~\ref{tab:flux_limits}. We report null results from 
Coma, Bullet, El Gordo, and SPT-CLJ 2012-5649. We calculate the 95\% c.l. upper flux limits on the differential  photon flux (at 30 MeV), integral photon flux, and energy flux. The energy flux limits  are $\sim 10^{-10}$ erg/$cm^2$/sec for all the clusters with  no detections (cf. Table~\ref{tab:flux_limits}).  Therefore, we do not confirm the gamma-ray line seen for the Coma cluster in I04.
For the  VIRGO cluster, we detect a signal between 0.75 and 1.50 MeV with significance between $2.5-2.8\sigma$ and null detections  at higher energies up to 30 MeV. The SED for the VIRGO cluster using all the three search templates can be found in Figs.~\ref{fig:Figure5},~\ref{fig:Figure6},~\ref{fig:Figure7}. The observed energy flux is between $(0.5-1.25) \times 10^{-9}~\rm{erg/cm^2/sec}$, depending on the template used. The observed signal is at a lower energy than that observed in  I04. Therefore, although we see a signal at low significance around 1~MeV, we do not corroborate  the gamma-ray line observed for VIRGO in I04.
\rthis{ We note that since M87 (in the center VIRGO cluster) contains a misaligned jet from which VHE emissions have been detected, this AGN in M87 could also give rise to the observed MeV emission. } 

Therefore, this pilot search for MeV gamma-ray emission with COMPTEL from a few selected  galaxy clusters is a proof of principles application of the revamped COMPTEL software in the analysis of galaxy clusters and paves the way for extension to other clusters. We shall also extend this analysis to  X-ray and SZ selected samples. \rthis{We shall also discuss theoretical implications of our results,  including constraints on inverse-Compton emission and possible limits on magnetic fields for each of the clusters in a future work.}
More detailed studies  of galaxy clusters at MeV energies should soon be possible with future missions such as COSI~\cite{COSI} or GECCO~\cite{GECCO}, which are sensitive up to 10 MeV.  Beyond 10 MeV, a whole bunch of experiments have been proposed but not yet scheduled for launch at the time of writing, such as GAMMA-400~\cite{Galper2014,Egorov2020}, Advanced Compton Telescope~\cite{Boggs2006}, Advanced Energetic Pair Telescope (AdEPT)~\cite{Hunter2014}, PANGU~\cite{Wu2014,Wu2015} , GRAMS~\cite{Aramaki2020,Aramaki2020b}, MAST~\cite{Dzhatdoev2019}, All-Sky-ASTROGAM~\cite{Tatischeff2019}. All of these missions should provide more insights into gamma-ray emission from galaxy clusters at energies greater than 10 MeV.

\rthis{In the spirit of open science, we have made our analyses templates and data used for making the SED plots publicly available at
\url{https://github.com/siddhantmannaiith/Comptel-Analysis-}}

\begin{acknowledgments}
SM extend his sincere gratitude to the Government of India, Ministry of Education (MOE) for their continuous support through the stipend, which has played a crucial role in the successful completion of our research.  We are also very grateful to 
J\"urgen Kn\"odlseder for  thoroughly explaining the  nuances of the COMPTEL analysis software and helping us   with the  analysis. \rthis{We also acknowledge the anonymous referee for several useful comments on the manuscript.}

\end{acknowledgments}

\bibliography{references}

\end{document}